\documentclass[prl,twocolumn,showpacs,superscriptaddress,preprintnumbers,amsmath,amssymb,floatfix]{revtex4}

\usepackage{bm}
\usepackage{graphicx}
\usepackage{amsbsy}
\usepackage{amsmath}
\usepackage{amsfonts}
\usepackage{amsthm}
\usepackage{color}
\usepackage{mathrsfs}

\usepackage{graphicx}
\usepackage{dcolumn}
\usepackage{bm}

\begin{document}


\title{Coherent control of low loss surface polaritons}

\author{Ali Kamli}
\email{akamli@qis.ucalgary.ca}
\affiliation{Institute for Quantum Information Science, University of Calgary,
	Calgary, Alberta, Canada T2N 1N4}
\affiliation{Department of Physics, King Khalid University, Abha  61314, P O Box 9003, Saudi Arabia}
\affiliation{The National Centre for Mathematics and Physics, KACST, Riyadh, Saudi Arabia}
\author{Sergey A. Moiseev}
\email{smoiseev@qis.ucalgary.ca}
\affiliation{Institute for Quantum Information Science, University of Calgary,
	Calgary, Alberta, Canada T2N 1N4}
\affiliation{Kazan Physical-Technical Institute of the Russian Academy of Sciences, 10/7 Sibirsky Trakt, Kazan, 420029, Russia}
\author{Barry C. Sanders}
\affiliation{Institute for Quantum Information Science, University of Calgary,
	Calgary, Alberta, Canada T2N 1N4}

\date{\today}


\begin{abstract}
We propose fast all-optical control of surface polaritons (SPs) by placing an electromagnetically
induced transparency (EIT) medium at an interface between two materials.
EIT provides longitudinal compression and a slow group velocity while matching
properties of the two materials at the interface provides strong transverse confinement.
In particular we show that an EIT medium near the interface between a dielectric and a negative-index
metamaterial can establish tight longitudinal and transverse confinement plus extreme
slowing of SPs, in both transverse electric and transverse magnetic
polarizations, while simultaneously avoiding losses.
\end{abstract}


\maketitle


\emph{Introduction:---}
All-optical rapid guidance, processing, and control of light in nanophotonic and quantum information
applications is important but limited by weak nonlinearities in typical materials, the fast speed of light,
and physical limitations to confinement such as enhanced absorption
and the narrow spectral width of light. Strategies to overcome these speed and confinement weaknesses
include exploiting photonic crystals with defect structures to trap light~\cite{JJ02},
surface polaritons (SPs) to confine light to wavelength dimensions~\cite{AM82} and
use nonlinear interactions~\cite{BIH91,Mil05},
and electromagnetically induced transparency (EIT) to slow and compress light in the longitudinal
propagation direction~\cite{HH99}, which are realized in solid state~\cite{TSSMHH02},
Bose-Einstein condensates~\cite{LDBH01}, and Mott insulators~\cite{CAB+08}.
We propose placing an EIT medium at the interface of two materials.
This arrangement benefits from the
combination of transverse confinement of surface polaritons with the longitudinal compression
and slowing of pulses due to EIT.

We derive an analytical solution to the problem, 
which provides an elegant picture of EIT at a material interface,
including hitherto unsuspected properties for the interface between a dielectric and
a negative-index meta material (NIMM)~\cite{EZ06}:
our two key results for the dielectric-NIMM interface are
(i)~that this interface simultaneously supports both transverse electric (TE)
and transverse magnetic (TM) polarizations whereas dielectric-dielectric
and dielectric-metal interfaces only supports TM, and 
(ii)~SP loss can be made arbitrarily
small for the dielectric-NIMM interface but not for the other cases.
Our analysis applies to general interfaces, but the dielectric-NIMM interface is especially
intriguing and suggests possibilities of low-loss, fast control of both TM and TE polarized SPs.

\begin{figure}[h]
\centering
  \includegraphics[width=80mm]{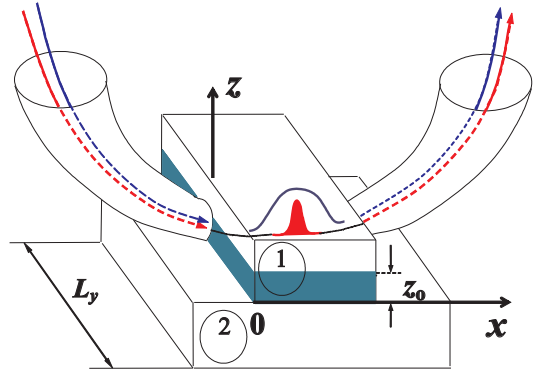}
  \caption{(Color online) End-fire coupling scheme for coherent control of surface polaritons (SPs).
 Two incoming beams from the left waveguide create two SP pulses above the
 interface between media~1 and~2 in half-spaces $z>0$ and $z<0$, respectively:
 control (blue line) and signal probe (red filled).
 The SPs interact with a collection
 of~$\Lambda$ medium shown above the interface as a shaded green layer of thickness $z_0$.}
\label{fig:scheme}
\end{figure}

For our scheme in Fig.~\ref{fig:scheme},
we suggest end-fire coupling excitation, which has been
demonstrated experimentally with a high efficiency of~$0.7$~\cite{Mai07}.
Our approach is to create two SP fields by directing two laser beams
at the interface between two materials with permittivities $\varepsilon_i$ and
permeabilities $\mu_i$ (top: $i=1$; bottom: $i=2$),
and these two SP fields propagate toward the EIT medium comprising three-level $\Lambda$ atoms (3LA),
quantum dots, nitrogen-valence centers in diamond, rare-earth ions in crystals, or similar (henceforth the `$\Lambda$ medium');
the three levels are designated $|\ell\rangle$ for $\ell\in\{1,2,3\}$, and the transition frequency
$\omega_{\ell\ell'}$ corresponds to $|\ell\rangle\leftrightarrow|\ell'\rangle$.
Our analysis does not restrict the signs of $\varepsilon$ and~$\mu$ hence
accommodates dielectrics, surface plasmons at a dielectric-metal interface,
and also NIMMs where both~$\varepsilon, \mu<0$.

\emph{Absorption and dispersion of SP fields:---}
The SP fields, which propagate
in the positive $x$-direction, are characterized by the $\omega$-dependent complex wave vector
$k_\parallel+\text{i}\kappa$.
For $k_j^2=k_\parallel^2-\omega^2\varepsilon_j\mu_j/c^2$,
the wave vector component normal to the interface,
$k_1\varsigma_2=-k_2\varsigma_1$ where $\varsigma\equiv\mu$ for the TE mode and
$\varsigma\equiv\varepsilon$ for the TM mode;
both polarizations coexist only if both conditions are simultaneously satisfied.

We analyze the SP modes at the interface
between a dielectric  and a NIMM media.
Although the dielectric-NIMM interface is technically challenging, rapid progress is
bringing metamaterials to the optical domain~\cite{CKYCXDS07}
thereby opening possibilities for optical storage and control~\cite{TBH07}.
The dielectric-NIMM interface is especially attractive
because both TE and TM polarization modes
can co-exist and, as we show below, complete suppression of SP losses is possible in principle
thereby admitting novel opportunities in SP field control.
Optical properties of the NIMM can be modeled with
complex dielectric permittivity and magnetic permeability given by~\cite{Mai07,EZ06}:
\begin{equation}
\label{eq:varsigma_2}
	\varsigma_2(\omega)
		=\varsigma_\text{r}+\text{i}\varsigma_\text{i}
		=1-\frac{\omega_\text{f}^2}{\omega(\omega+\text{i}\gamma_\text{f})}
\end{equation}
for the two cases $\varsigma\equiv\varepsilon$, $\text{f}\equiv\text{e}$ and
$\varsigma\equiv\mu$, $\text{f}\equiv\text{m}$.
Here~$\omega_{\text{e},\text{m}}$ are the electric and magnetic plasma frequencies of the NIMM
and $\gamma_{\text{e},\text{m}}$ the loss rates, respectively.
Accounting for complex~$\varepsilon$ and~$\mu$  in the surface boundary conditions of SP wave vectors,
we find (TM case)
\begin{equation}
\label{eq:permperm}
	k_\parallel(\omega)+\text{i}\kappa(\omega)
		=\frac{\omega}{c}\sqrt{\varepsilon_1\varepsilon_2(\varepsilon_2\mu_1-\epsilon_1\mu_2)/
			(\varepsilon_2^2-\varepsilon_1^2)},
\end{equation}
and the TE case is similar.
The real (imaginary) part of Eq.~(\ref{eq:permperm}) yields SP TM mode
dispersion (loss).

We study the system numerically at room temperature and at optical frequencies to determine its features.
For metal, we use the values for Ag~\cite{OBA+85}:
$\omega_\text{e}=1.37\times10^{16}\text{s}^{-1}$,
$\gamma_\text{e}= 2.73 \times10^{13}s^{-1}$ and $\varepsilon_{1}=1.3$, $\mu_{1}=1$.
As NIMM technology is embryonic, we consider a wide range of magnetic plasmon
frequency such that $\omega_\text{m}\leq0.5 \omega_\text{e}$~\cite{Mai07},
and $\gamma_\text{m}$ is between $10^{-5}\gamma_\text{e}$ and $\gamma_\text{e}$ itself.

\begin{figure}[h]
  \centering
 \includegraphics[width=80mm]{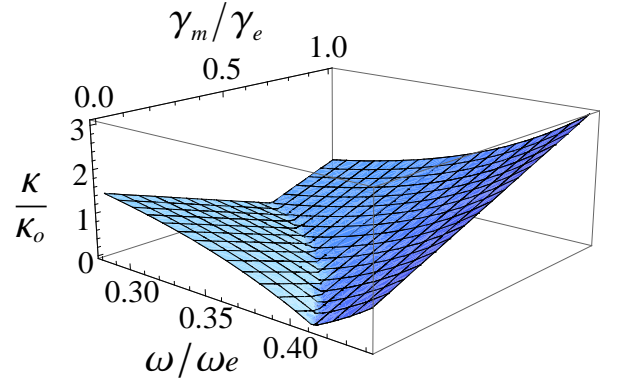}
 \caption{(Color online) Absorption loss for surface polaritons as a function of frequency ~$\omega/\omega_\text{e}$
and magnetic decoherence rate $\gamma_\text{m}/\gamma_\text{e}$ , where $\kappa_{0}=10^4\text{m}^{-1}$, $\gamma_\text{e}= 2.73 \times10^{13}\text{s}^{-1}$, $\omega_\text{m}=0.5 \omega_\text{e}$ and $\varepsilon_{1}=1.3$, $\mu_{1}=1$.}
\label{fig:Absorption}
\end{figure}

\begin{figure}[h]
  \centering
  \includegraphics[width=80mm]{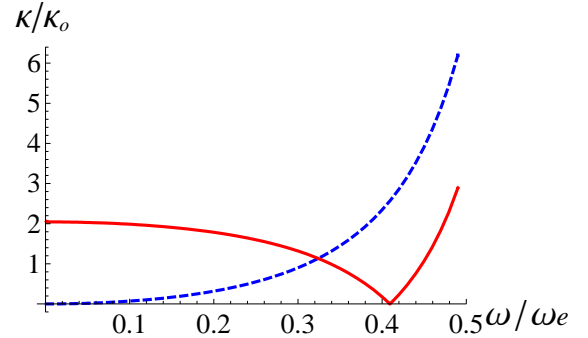}
  \caption{(Color online) Comparison of SP losses at a dielectric-NIMM interface (red curve) and losses of surface plasmons at
a dielectric-metal interface (blue curve) for $\gamma_\text{m}=\text{10}^{11}\text{s}^{-1}$
, $\gamma_\text{e}= 2.73 \times10^{13}\text{s}^{-1}$ and $\omega_\text{m}=0.5\omega_{e}$,
$\omega_\text{e}=1.37\times10^{16} \text{s}^{-1}$, $\omega_0 \cong0.4092\omega_\text{e}$.}
\label{fig:kappa2d}
\end{figure}

Numerical results for~$\kappa(\omega)$ are shown in Fig.~\ref{fig:Absorption}
and reveal a deep abyss for all~$\omega$; 
the abyss frequency~$\omega_0$
corresponds to a specific ratio of magnetic and electric loss for each~$\omega$,
and Fig.~\ref{fig:Absorption} reveals $\kappa(\omega_0)\sim0$, i.e.\ a complete cancelation of losses.
From Eq.~(\ref{eq:permperm}) $\omega_0$ is determined with high accuracy from
\begin{equation}
\label{eq:cond}
	\frac{\mu_\text{i}}{\varepsilon_\text{i}}
		=\frac{\mu_\text{r}(\varepsilon_\text{r}^2+\varepsilon_\text{1}^2)
			-2\varepsilon_\text{r}\varepsilon_\text{1}}{\varepsilon_\text{r}
			(\varepsilon_\text{r}^2-\varepsilon_\text{1}^2)},
\end{equation}
which cannot be satisfied for an interface  
between a dielectric and metal because $\mu_1,\mu_2>0 $.
Figure~\ref{fig:kappa2d} compares SP losses at a dielectric-NIMM interface for
$\gamma_\text{m}=\text{10}^{11}\text{s}^{-1}$ and $\omega_\text{m}=0.5\omega_\text{p}$ and surface plasmon polaritons at a dielectric-metal interface.

We observe that NIMMs enable absorption losses to be drastically reduced in a narrow frequency band,
as studied for freely propagating fields~\cite{KAS07}.
Here we predict a similar effect but for SP fields using a typical NIMM interface.
Equation~(\ref{eq:cond}) shows that $\varepsilon_\text{r},\mu_\text{r}<0$  implies
destructive interference between electric and magnetic absorption responses,
which explains the high suppression of losses around  $\omega_0$.
The relative weakness of decay rates makes the frequency $\omega_0$ sensitive to electric and magnetic decoherence rates as is evident from Fig.~(\ref{fig:Absorption}).
Henceforth we demonstrate the possibility of coherent control of slow SP within frequency range around
$\omega_0$ of complete reduction of their losses in basic materials.

\emph{EIT control of SP modes:---}
Here we give a general analysis of EIT based coherent control of  SP pulse interacting with a $\Lambda$ medium near the surface.
We assume that the probe field in the TM mode has a frequency~$\omega_{31}$
and control field frequency equal to~$\omega_{32}$.
The transition $|1\rangle\leftrightarrow|2\rangle$ is dipole-forbidden.

The SP electric field near the surface is obtained from field quantization~\cite{AB00} in a dispersive medium~\cite{LLP06}.
In the plane wave expansion over modes indexed by~$\lambda$,
\begin{equation}
\label{eq:E}
	\hat{\bm{E}}(x,z)=\sum_\lambda\int\text{d}k_\parallel
	\left[\bm{E}_{0\lambda} \left(k_\parallel,z\right)\hat{a}_\lambda\left(k_\parallel\right)\text{e}^{\text{i}k_\parallel x}
		+\text{hc}\right],
\end{equation}
with
\begin{equation}
	\left[\hat{a}_\lambda\left(k_\parallel\right),\hat{a}^\dagger_{\lambda'}(k'_\parallel)\right]
		=2\pi\delta_{\lambda\lambda'}\delta\left(k_\parallel-k'_\parallel\right).
\end{equation}
We develop the theory for TM; the TE case is then a straightforward generalization.
Dropping~$\lambda$, we obtain
\begin{equation}
\label{eq:E0}
	\bm{E}_0\left(k_\parallel,z\right)
		= \left\{\begin{array}{ll}
			\left(\bm{e}_{x}+\text{i} \bm{e}_{z} k_\parallel/k_1\right)
				E_0\left(k_\parallel\right)\text{e}^{-k_1z},	&z>0,	\\
			\left(\bm{e}_{x}-\text{i}\bm{e}_{z} k_\parallel/k_2\right)
				E_0\left(k_\parallel\right)\text{e}^{k_2z},&z<0\end{array}\right. .
\end{equation}
Here $\bm{e}_{x},\bm{e}_{z}$ are unit vectors along the $\bm{x},\bm{z}$ directions,
with electric field amplitude
\begin{equation}
\label{eq:Nkparallel}
	E_0\left(k_\parallel\right)=\sqrt{\hbar\omega\left(k_\parallel\right)/
		2\pi\varepsilon_oL_yL_z(\omega,\varepsilon,\mu)},
\end{equation}
and transverse quantization length
$L_z=D+\frac{\omega^2\left(k_\parallel\right)}{c^2}S$ with
\begin{align}
\label{eq:DS}
	D=&\frac{\partial}{\partial\omega}(\omega\varepsilon_1)\frac{k_1^2+k_\parallel^2}{k_1^3}
		+\frac{\partial}{\partial\omega}(\omega\varepsilon_2)\frac{k_2^2+k_\parallel^2}{k_2^3},
			\nonumber	\\
	S=&\frac{\partial}{\partial\omega}(\omega\mu_1)\frac{\varepsilon_1^2}{k_1^3}
		+\frac{\partial}{\partial\omega}(\omega\mu_2)\frac{\varepsilon_2^2}{k_2^3}.
\end{align}
These quantities depend on the interface parameters and on the SP mode dispersion relation
$\omega\left(k_\parallel\right)$ of Eq.~(\ref{eq:permperm}).

Adopting the usual EIT approximations~\cite{Mil05,FIM05}
for the evolution of a SP interacting with a $\Lambda$ medium,
we find the Fourier SP field equation
\begin{align}
\label{eq:A}
	(\partial/{\partial x}-\text{i}{\nu}/{v_0})\hat{A}(\nu,x)
		=-\left[\alpha(\nu)+\kappa(\omega_{31})\right]\hat{A}(\nu,x),
\end{align}
on the surface:
$\hat{A}(\nu,x)=(2\pi)^{-1}\int\text{d}t\text{e}^{\text{i}\nu t}\hat{A}(t,x,z=0)$,
where $\hat{A}(t,x,z=0)=\int\text{d}k_\parallel E_{0} \left( k_\parallel \right)\hat{a}(k_\parallel,t)\text{e}^{\text{i}k_\parallel^s x}$.
By solving Eq.~(\ref{eq:A}) we obtain the electric field over the surface
\begin{align}
\label{eq:solution of SP}
	\hat{\bm{A}}(t,x,z>0)
	= \left(\bm{e}_{x}+\text{i} \bm{e}_{z} k_\parallel(\omega_{31})/k_1^{s}\right)
		e^{-k_1^{s}z}\nonumber	\\
		\int \text{d}\nu\,
		e^{-\text{i}\nu t+[\text{i}\frac{\nu}{v_0}
		-\alpha(\nu)-\kappa(\omega_{31})]x}
		\hat{A}(\nu,0)
\end{align}
where $\nu\equiv\omega\left(k_\parallel\right)-\omega_{31}$
is the SP field detuning from the central frequency~$\omega_{31}$,
which is assumed to be close to $\omega_0$.
Here $v_0 =\partial \omega/\partial k_\parallel$  is the SP group velocity without a 3LA $\Lambda$ medium at $\omega=\omega_{31}$, and
\begin{equation}
\label{eq:alpha}
	\frac{\alpha(\nu)}{2\pi}=\frac{|g|^2}{v_0}
		\int_0^\infty\!\!\!\int_0^{L_y}\text{d}y\text{d}z\,
		\frac{n(\bm{r})(\gamma_{21}-\text{i}\nu)\text{e}^{-2k_1^\text{s} z}}
		{\left|\Omega_\text{c}(\bm{r})\right|^2-\left(\nu+\text{i}\gamma_{21}\right)
			\left(\nu+\text{i}\Gamma_{31}\right)}
\end{equation}
yields dispersion and absorption of the SP field,
with~$\Gamma_{31}$ a linewidth,
$\Omega_\text{c}(\bm{r})$ the control field Rabi frequency,
$g=\textbf{d}\cdot\left(\bm{e}_{x}+\text{i} \bm{e}_{z} k_\parallel/k_1\right)
E_{0} \left( k_\parallel(\omega_{31}) \right) / \hbar$
the SP-$\Lambda$ coupling constant, $\textbf{d}$ the atomic dipole moment,
and $n(\bm{r})$ the $\Lambda$ medium density.
These parameters can be optimized for SP field control. Also the control field, which yields~$\Omega_\text{c}$,
can be a freely propagating mode or a SP TE or TM field.
Henceforth we assume $|\Omega_\text{c}(\bm{r})|^2=|\Omega|^2\text{e}^{-2k_1^c z}$,
where  $k_1^\text{s}$ and $k_1^\text{c}$ are the probe and control wave vectors in the $z$-direction for medium~1 as in Eq.~(\ref{eq:E0}).

Absorption and dispersion of the SP-field in the presence of a 3LA medium is given by
 $\alpha(\nu)+\kappa(\omega_{31})$ with inhomogeneous broadening at resonance transition
$h=\Delta_\text{w}/\pi(\Delta^2+\Delta_\text{w}^2)$ and inhomogeneous broadening width $\Delta_\text{w}$ .
Thus $\Gamma_{31}=\Delta_\text{w}+\gamma_{31}$
where we have assumed that all atoms share the same decay constants~$\gamma_{mn}$;
this assumption is reasonable for solid-state systems at liquid He temperatures
because~$\gamma_{31}/\Delta_\text{w}$ is negligible~\cite{Mil05}.
From Eq.~(\ref{eq:alpha}), it follows that
the field amplitude effectively is bounded by
$z \leq \text{min}\{1/k_1^\text{s},1/k_1^\text{c}\}$, and we assume confinement
of the $\Lambda$ medium to this height above the interface.

Eq.~(\ref{eq:alpha}) is integrable for constant density ~$n(\bm{r})\equiv n(0<z<z_o)=n$, yields
\begin{equation}
	\alpha(\nu,z_0)=\alpha_0(\omega_{31})
		G\left(k_1^\text{s},k_1^\text{c},z_0,\beta(\nu)\right)
\end{equation}
with
\begin{align}
\label{eq:G}
	G(k_1^\text{s},&k_1^\text{c},z_0,\beta)
		=\frac{\text{i}\Gamma_{31}}{\nu+\text{i}\Gamma_{31}}
		\Bigg[  {_2F_1}\left(1,\frac{k_1^\text{s}}{k_1^\text{c}},\frac{k_1^\text{s}+k_1^\text{c}}{k_1^\text{c}},
			\frac{1}{\beta(\nu)}\right)
					\nonumber \\	&
		-\text{e}^{-2k_1^\text{s}z_0}\; _2F_1\left(1,\frac{k_1^\text{s}}{k_1^\text{c}},
			\frac{k_1^\text{s}+k_1^\text{c}}{k_1^\text{c}},
			\frac{\text{e}^{-2k_1^\text{c}z_0}}{\beta(\nu)}\right)\Bigg]
\end{align}
a spectral function,
$_2F_1$ the hypergeometric function, and
$\beta
		\equiv(\nu+\text{i}\gamma_{21})
		(\nu+\text{i}\Gamma_{31})/|\Omega|^2$.
Function~(\ref{eq:G}) is a maximum $G=1$ for~$\nu=0$
and depends on $\Delta_\text{w}$, $k_1^\text{s}/k_1^\text{c}$, $z_0$, and~$\Omega_\text{c}$,
which provides rich opportunities for spectral and spatial control of the SP field.

As the formula is integrable, the resonant absorption coefficient for the $\Lambda$ medium at $\omega_{31}$
is expressed in a simple form if the control field is off and $z_0\gg 1/k_1^\text{s}$:
\begin{equation}
\label{eq:resonant absorption coefficicent}
	\alpha_0(\omega_{31})=\pi n L_y|g|^2/k_1^\text{s}v_0(\omega_{31})\Gamma_{31}
\end{equation}
with coupling constant $|g|^2 \sim 1/L_z$.
An appropriate choice of materials can lead to small~$L_z$ hence considerably enhance the SP field amplitude and   increase interaction coupling between the SP field and the 3LA $\Lambda$ medium.

\emph{Numerical example:---}
Using our solution we demonstrate the exciting possibility of EIT control for the \emph{low loss SP modes},
which opens opportunities to exploit spatial confinement and temporal control of SPs
via their interaction with the $\Lambda$ medium.

For
$L_y = 2.5\mu m,\;0.4087 \omega_{e}<\omega_{31}<0.4097 \omega_{e}$,
we find
$\kappa (\omega_{31})<0.01 \kappa_{0},\;v_0(\omega_{31})\approx0.6 c,\;
k_1^\text{s}(\omega_{31})\approx 1 \mu m^{-1}$.
From Eqs.~(\ref{eq:Nkparallel}) and~(\ref{eq:resonant absorption coefficicent})
for resonant optical transitions of rare earth ions in crystals, e.g.\ Pr$^{+3}$-doped Y$_2$SiO$_5$ 
(demonstrated for EIT experiments~\cite{TSSMHH02})
with density $n\approx 10^{24}\text{m}^{-3}$, $\Gamma_{31}\cong10^9\text{rad}/\text{s}$, 
$\gamma_{21}^{-1}\gg 1 \mu s$, $\varepsilon_1 \cong 1.3$,  and
$|\textbf{d}| \approx - e a_0$ (with~$e$ the electron charge and $a_0$ the Bohr radius), 
we find $\alpha_0(\omega_{31})\approx 10\mu m^{-1}$.
Using Eq.~(\ref{eq:solution of SP}) we compute time delay and group velocity for
a Gaussian amplitude envelope of the SP probe input pulse
$\exp\left[-(t/\delta t)^2/2\right]$ in the medium at $x=0$.

\begin{figure}[h]
  \centering
  \includegraphics[width=80mm]{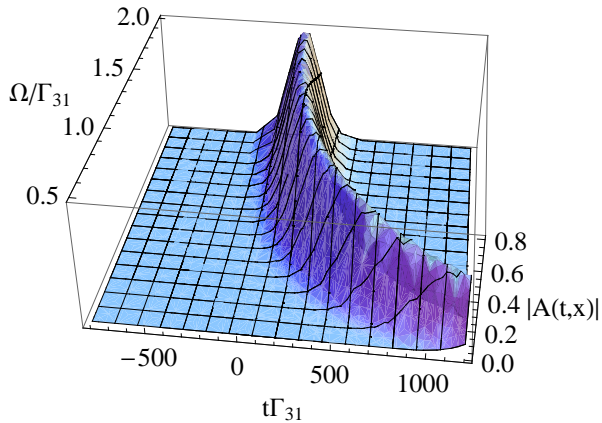}
  \caption{(Color online) 3D graph of pulse propagation profile as a function of time $t\Gamma_{31}$
  and control field amplitude $\Omega/\Gamma_\text{31}$, $\Gamma_{31}=10^9\text{rad}/\text{s}$,
  $\kappa (\omega_{31})=0.01 \kappa_{0}= 0.1 \times10 ^{3} m^{-1}$.}
\label{fig:pulseA}
\end{figure}

\begin{figure}[h]
  \centering
  \includegraphics[width=80mm]{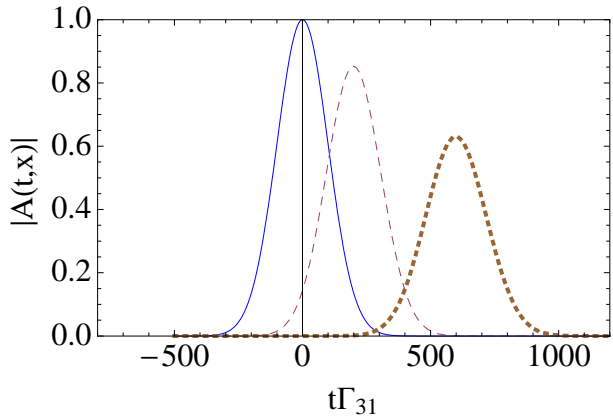}
  \caption{(Color online) The pulse propagation profile as a function of time, $t\Gamma_{31}$,
  at different locations near the interface: blue (solid) at $x=0$, red (dashed) at $x_1 =1$mm,
  and brown (dotted) at $x_2=3$mm.}
\label{fig:pulseB}
\end{figure}

Input and output pulse profiles are presented in Figs.~\ref{fig:pulseA},~\ref{fig:pulseB} for input pulse duration
$\delta t=100$ns, for media lengths $x_1=1$mm and $x_2=3$mm ($<L=1/\kappa$).
As seen in Fig.~\ref{fig:pulseA} the time delay decreases with the control field amplitude as $\Omega^{-2}$. Fig.~\ref{fig:pulseB} shows the pulse profile for
 $\Omega=\Gamma_\text{31}$ when it propagates a distance $x_1=1$mm and $x_2=3$mm.
The time delays are
$t_\text{delay}=2\delta t, 6\delta t$ respectively,
whereas the amplitude has decreased only by factors 0.85 and 0.65, respectively.
Thus the propagation length increased by more than
$500/\alpha_0(\omega_{31})$
due to EIT of the SP field.
Using these results we estimate the group velocity
$v_\text{g}\approx 5000\text{m}/\text{s}$ and a compressed longitudinal envelope of
the SP pulse
$l_\text{SP}=v_\text{g}\delta t=0.5\text{mm}$, i.e.\
much smaller than the medium size.
Thus the SP pulse can be successfully stored in the long-lived atomic coherence
$\rho_{12}(t,x)$
and subsequently retrieved in accordance with an EIT quantum memory protocol~\cite{FIM05}.

\emph{Conclusion:--}
We have demonstrated low loss SPs at the interface between two media
with quite general electromagnetic properties, including dielectrics, metals and
metamaterials, and derived a closed form solution that provides deep insight into
the system and control of spatially confined slow SP fields.
We show that light pulses can be stored at the interface of two media exploiting EIT
and low loss SP fields near a NIMM-dielectric interface.

\acknowledgments
We gratefully acknowledge financial support from $i$CORE, NSERC, CIFAR,
KACST (Saudi Arabia), and RFBR grant \# \!06-02-16822.

\end{document}